# Electronic structure of above-room-temperature van der Waals ferromagnet Fe$_3$GaTe$_2$

Ji-Eun Lee[1,2†], Shaohua Yan[3,4†], Sehoon Oh[5], Jinwoong Hwang[6], Jonathan D. Denlinger[1], Choongyu Hwang[7,8], Hechang Lei[3,4*], Sung-Kwan Mo[1*], Se Young Park[5*], Hyejin Ryu[9*]

[1]Advanced Light Source, Lawrence Berkeley National Laboratory, Berkeley, CA 94720, USA

[2]Max Planck POSTECH Center for Complex Phase Materials, Pohang University of Science and Technology, Pohang 37673, Korea

[3]Beiing Key Laboratory of Optoelectronic Functional Materials MicroNano Devices, Department of Physics, Renmin University of China, Beijing 100872, China

[4]Key Laboratory of Quantum State Construction and Manipulation (Ministry of Education), Renmin University of China, Beijing, 100872, China

[5]Department of Physics and Origin of Matter and Evolution of Galaxies (OMEG) Institute, Soongsil University, Seoul 06978, Korea

[6]Department of Physics, Kangwon National University, Chuncheon 24341, Korea

[7]Department of Physics, Pusan National University, Busan 46241, Korea

[8]Quantum Matter Core-Facility, Pusan National University, Busan 46241, Korea

[9]Center for Spintronics, Korea Institute of Science and Technology (KIST), Seoul 02792, Korea

† These authors contributed equally to this work.

**Abstract**

Fe$_3$GaTe$_2$, a recently discovered van der Waals (vdW) ferromagnet, demonstrates intrinsic ferromagnetism above room temperature, necessitating a comprehensive investigation into the microscopic origins of its high Curie temperature ($T_C$). In this study, we reveal the electronic structure of Fe$_3$GaTe$_2$ in its ferromagnetic ground state using angle-resolved photoemission spectroscopy (ARPES) and density-functional theory (DFT) calculations. Our results establish a consistent correspondence between the measured band structure and theoretical calculations, underscoring the significant contributions of the Heisenberg exchange interaction ($J_{ex}$) and magnetic anisotropy energy (MAE) to the development of the high $T_C$ ferromagnetic ordering in Fe$_3$GaTe$_2$. Intriguingly, we observe substantial modifications to these crucial driving factors through doping, which we attribute to alterations in multiple spin-splitting bands near the Fermi level. These findings provide valuable insights into the underlying electronic structure and its correlation with the emergence of high $T_C$ ferromagnetic ordering in Fe$_3$GaTe$_2$.

The discovery of intrinsic ferromagnetic order in low-dimensional systems has provided a valuable platform for investigating fundamental magnetic phenomena and advancing the development of spintronic devices. Van der Waals (vdW) ferromagnets exhibit a range of intriguing magnetic properties, such as large magnetic anisotropy, the quantum anomalous Hall effect, strongly correlated spin systems, and long-range magnon excitations[1–6]. Moreover, their persistent magnetism down to monolayer presents promising opportunities for constructing innovative spin-based devices, including spin field-effect transistors and magnetoresistance memories[7–9]. Finding ferromagnetic materials that retain their magnetic properties at or above room temperature, such as $CrTe_2$ and $Fe_{5-x}GeTe_2$, holds significant implications for spintronics device applications[10–18].

Recently, a record-high $T_C$ of approximately 350-380 K has been discovered in single crystal $Fe_3GaTe_2$, a vdW layered ferromagnets[19], which is much higher than that of much studied its sister material $Fe_3GeTe_2$[20]. It also exhibits notable magnetic properties, including substantial perpendicular magnetocrystalline anisotropy, high saturation magnetic moment, and a large anomalous Hall angle, all persisting above room temperature. Consequently, this unprecedentedly high $T_C$ of $Fe_3GaTe_2$ has sparked immense interest, resulting in extensive research efforts primarily focused on investigating its basic magnetic properties and exploring spin-dependent electron transport in $Fe_3GaTe_2$ based devices[19,21–24]. However, the underlying microscopic mechanism responsible for the high $T_C$ ferromagnetic behavior in $Fe_3GaTe_2$ remains elusive. In particular, the electronic structure in its ferromagnetic ground state has not been carefully investigated, which would provide crucial information to delineate the similarities and differences in $Fe_3GaTe_2$ and $Fe_3GeTe_2$[2,3,5,25] systems to reveal the key physical parameters that enable high-temperature two-dimensional (2D) ferromagnetic order.

In this study, we report the electronic structure of $Fe_3GaTe_2$ single crystal in its ferromagnetic ground state using angle-resolved photoemission spectroscopy (ARPES) and first-principles density functional theory (DFT) calculations. Our experimental results of band structure demonstrate a good overall agreement with our DFT calculations. There are notable changes in both experimental and calculated low-energy electronic structures of $Fe_3GaTe_2$ compared with $Fe_3GeTe_2$[2,3,5,25] due to the changes in the valence of Ga and Ge, although both materials share similar characteristics in a large energy window. To understand the mechanism of high $T_C$ 2D ferromagnetism deeper, we delve into the fundamental driving factors based on our ARPES results, focusing particularly on the contributions of the Heisenberg exchange-coupling interaction ($J_{ex}$) and the magnetic anisotropy energy (MAE). Intriguingly, doping significantly influences these physical parameters, resulting in a reversal of the MAE direction and the emergence of in-plane magnetic ordering upon hole-doping. Our findings contribute to a comprehensive understanding of the electronic structure and ferromagnetic mechanism in $Fe_3GaTe_2$ while also suggesting the potential control of magnetic interactions through doping.

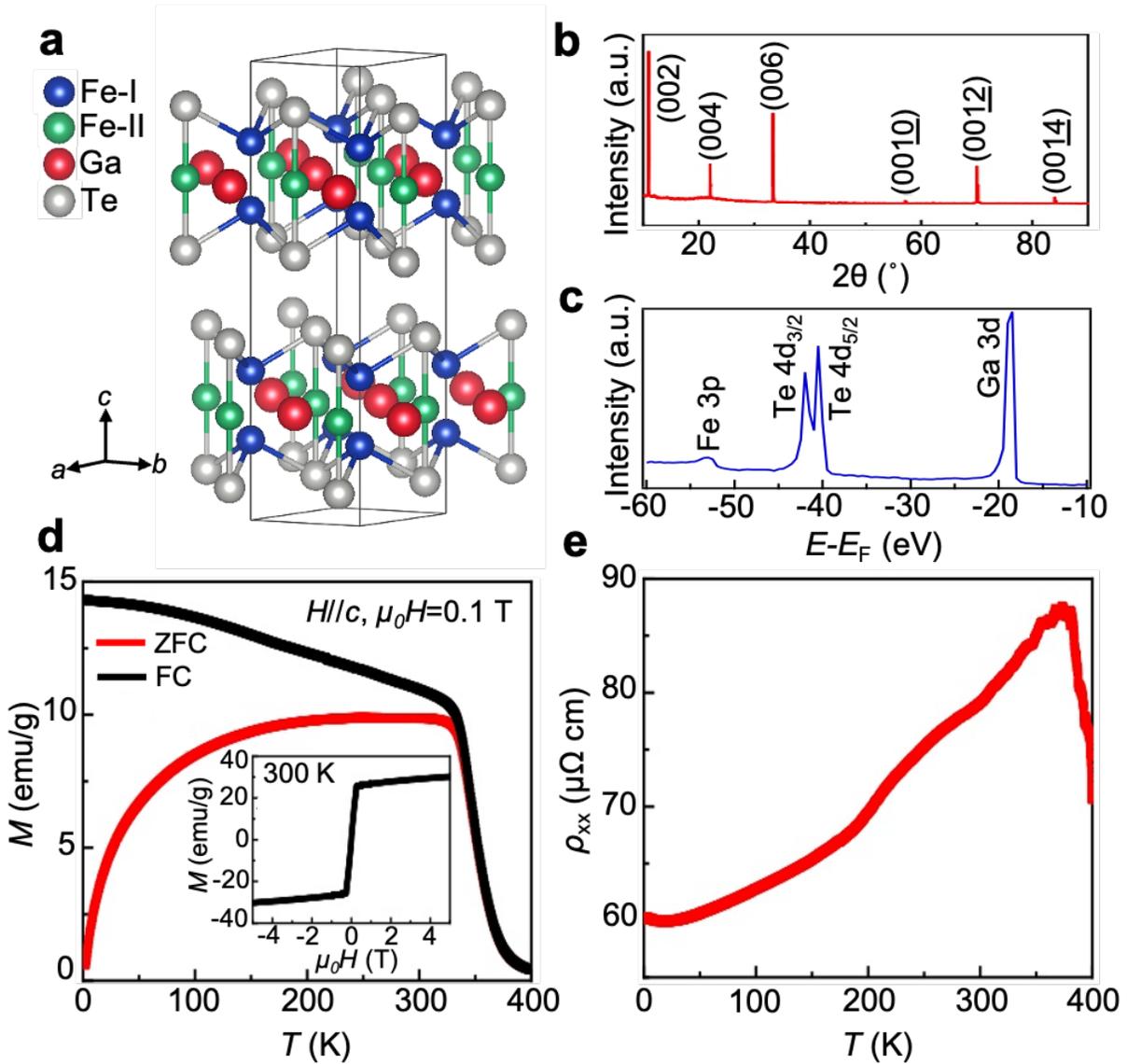

**Figure 1. Above room-temperature van der Waals layered ferromagnet Fe$_3$GaTe$_2$.** (a) Crystal structure of Fe$_3$GaTe$_2$. The rectangular box denotes the crystal unit cell. (b) The x-ray diffraction (XRD) patterns of Fe$_3$GaTe$_2$ single crystal. (c) Core level spectrum of bulk single crystal Fe$_3$GaTe$_2$, showing Fe 3$p$, Te 4$d$, and Ga 3$d$ peaks. (d) Temperature-dependent out-of-plane magnetization of Fe$_3$GaTe$_2$, measured under zero-field cooling (ZFC) and field cooling (FC). Inset shows magnetic moment as a function of the magnetic field at 300 K. (e) Temperature-dependent resistivity of Fe$_3$GaTe$_2$ from 2 K to 400 K.

$Fe_3GaTe_2$ is in a layered hexagonal crystal structure of a space group of $P6_3/mmc$ (No. 194), identical to that of well-known $Fe_3GeTe_2$[19], schematically shown in Fig. 1 (a). The Fe atoms are located in two inequivalent sites, denoted as Fe-I and Fe-II with different magnetic moments. $Fe_3Ga$ layers are sandwiched by the vdW-bonded Te layers on top and beneath. The slabs of $Fe_3GaTe_2$ are stacked along the $c$-axis with the weak vdW interaction. The x-ray diffraction (XRD) analysis confirms the single-crystalline nature of the synthesized $Fe_3GaTe_2$ without any impurities (Fig. 1 (b))[19]. The lattice constants we obtained from the XRD results are $a = b = 4.09(2)$ (Å) and $c = 16.07(2)$ (Å). The core level spectrum exhibits characteristic peaks of Fe $3p$, Te $4d$, and Ga $3d$ (Fig. 1 (c)). The temperature dependence of magnetization of $Fe_3GaTe_2$ single crystals exhibits a typical ferromagnetic behavior with $T_C \sim 380$ K (Fig. 1(d))[19], and the $M(\mu_0H)$ at 300 K also confirms the hard ferromagnetism of $Fe_3GaTe_2$ when $H//c$ (inset of Fig. 1(d)). In addition, it can be seen that $Fe_3GaTe_2$ single crystal shows a metallic behavior at low temperatures with an inflection point near $T_C$, which could be due to the suppression of spin-disorder scattering. All of these results are consistent with the results of the previous report[19].

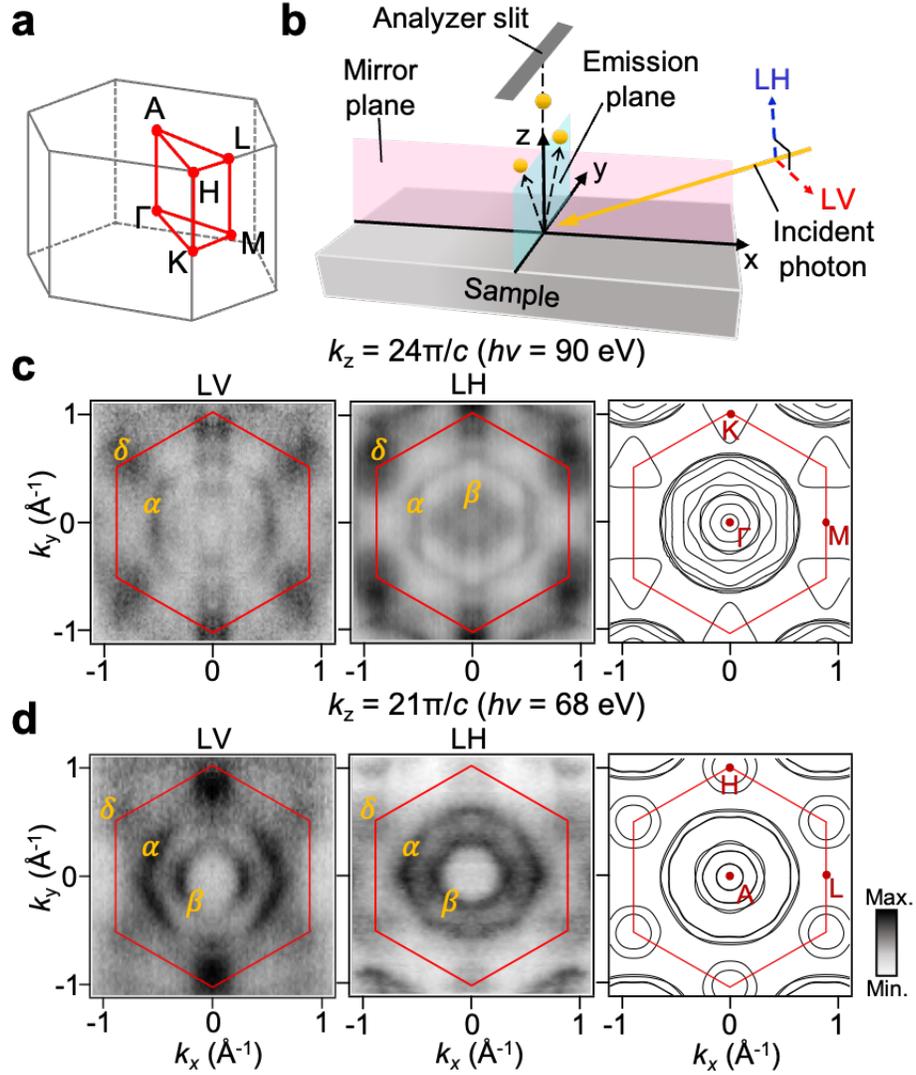

**Figure 2. The Fermi surface map of the ferromagnetic Fe$_3$GaTe$_2$.** (a) Three-dimensional (3D) Brillouin zone (BZ) of Fe$_3$GaTe$_2$ showing high symmetry points. (b) Schematic illustration of polarization geometry of the ARPES measurement. Linear horizontal (LH) and linear vertical (LV)-polarizations of light are defined with respect to the synchrotron light source. The analyzer slit is perpendicular to the mirror plane and parallel to the emission plane. (c), (d) Fermi-surface maps at $k_z = 24\pi/c$ and $21\pi/c$, respectively, corresponding to the photon energies of 90 and 68 eV, acquired using LV and LH polarized photons. The band notations ($\delta$, $\alpha$, and $\beta$) are labeled to describe the distinctive band features. The rightmost panels in (c) and (d) display the Fermi surface calculated by DFT. The region enclosed by the red hexagon represents the first Brillouin zone (BZ).

In order to investigate the nature of the magnetic properties of $Fe_3GaTe_2$, we performed ARPES measurements and DFT calculations on $Fe_3GaTe_2$ in its ferromagnetic ground state at 10 K (Figs. 2 and 3). By utilizing synchrotron-based ARPES measurements, we were able to tune the energy and polarization of photons, allowing us to capture selective band orbitals at different high symmetry points along the Γ-A direction. Despite the weak interlayer coupling characteristics of vdW layered materials, we observe a significant $k_z$ dispersion, comparable to that of $Fe_3GeTe_2$ (Figs. S1 in SI)[25]. However, we did not find a clear periodicity in the $k_z$ dispersion, also similar to the case of $Fe_3GeTe_2$, due to the large lattice constant along the c-direction that makes the periodicity along $k_z$ comparable to the $k_z$ broadening in the photoemission process, as well as strong photon energy dependence of photoemission cross-section of different orbitals. However, by comparing the orientation of the main Fermi surface (FS) around the center of the BZ with the results of DFT calculations, we were able to unambiguously define two photon energies corresponding to high-symmetry points, Γ and A. Figs. 2c and 2d show the FS maps obtained using both linear vertical (LV) and linear horizontal (LH) polarized lights at photon energies of 90 eV ($k_z = 24\pi/c$, Γ point, Fig. 2c) and 68 eV ($k_z = 21\pi/c$, A point, Fig. 2d), along with the corresponding calculated FS. At $k_z = 24\pi/c$, we observe multiple hole pockets (labeled as α and β bands) around the Γ point, with one of the outer pockets (α band) exhibiting a hexagonal shape aligned with the same orientation of Brillouin zone (BZ). There also exist triangular electron pockets around the K points (δ band). On the other hand, at $k_z = 21\pi/c$, we observe distinct double hole pockets (α and β bands), with the outermost hexagonal pocket (α band) rotated by 30 degrees relative to the 2D BZ, along with two circular electron pockets around the K points (δ bands). Both 30 degrees rotation of the hexagonal pocket (α and β bands) and the change in the number of electron pockets around K points (δ band). The calculated FS at the corresponding $k_z$ values exhibits the same characteristic features, demonstrating the presence of hole and electron pockets around the Γ and K points, respectively, as well as the rotation of the outer hexagonal hole pockets (α and β bands) as $k_z$ varies.

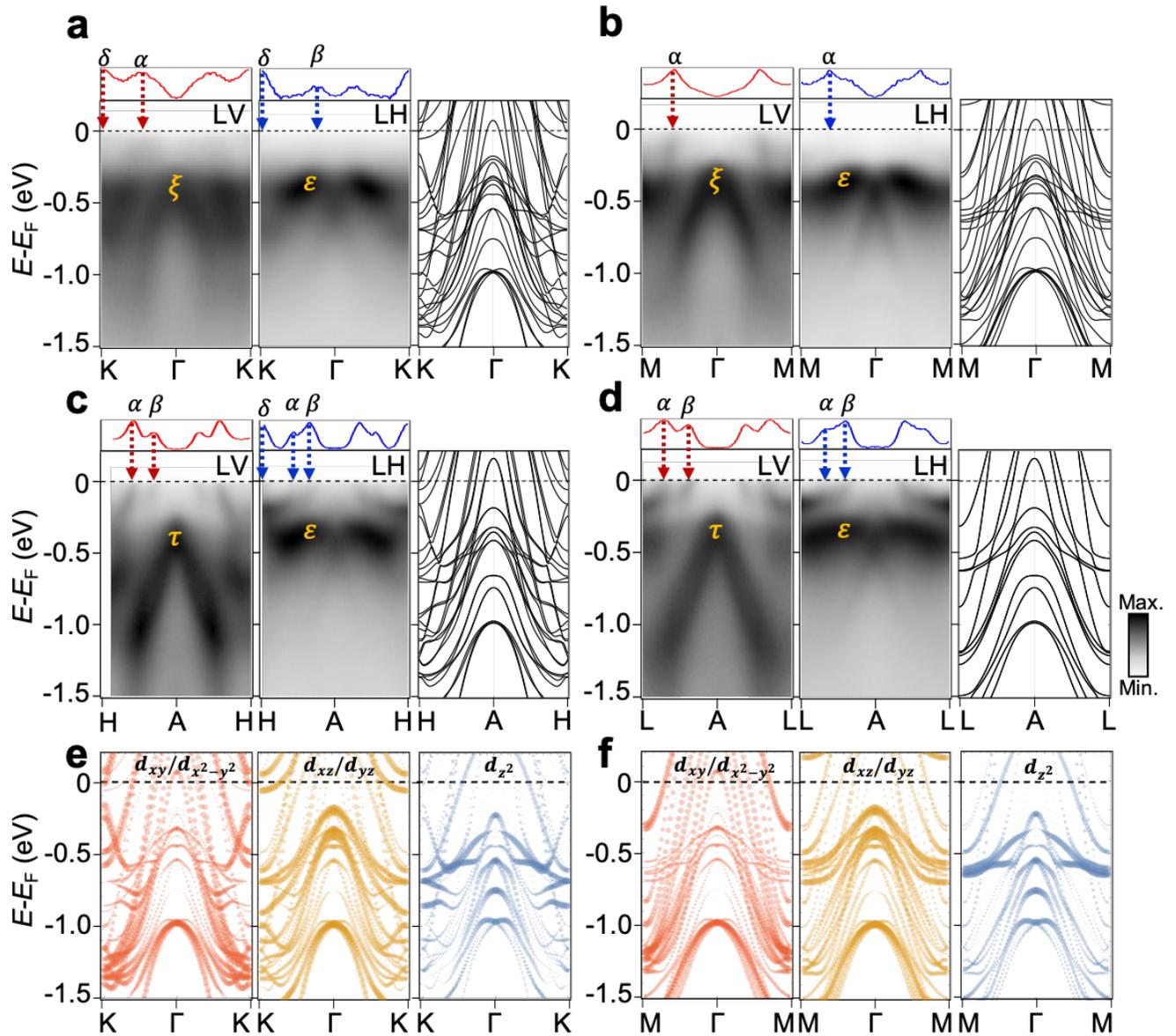

**Figure 3. Polarization-dependent ARPES band structures with orbital-resolved band calculation of Fe$_3$GaTe$_2$.** (a)-(d) The ARPES band dispersions along high-symmetry lines: (a) Γ-K, (b) Γ-M, (c) A-H, and (d) A-L (two left panels), along with corresponding DFT band structures

(rightmost panels). The upper panels show momentum distribution curves (MDCs) at $E-E_F$ = -50 meV, providing the band features near the Fermi level. The orbital-resolved band dispersions along the high-symmetry lines (e) Γ-K and (f) Γ-M are obtained by DFT projected with the $d_{xy}/d_{x^2-y^2}$, $d_{xz}/d_{yz}$, and $d_{z^2}$ orbitals from Fe-I and Fe-II atoms. The orbitals are grouped according to the point group $C_{3v}$ and $D_{3h}$ of the Fe-I and Fe-II atomic sites, respectively.

To gain a deeper understanding of the electronic structure of Fe$_3$GaTe$_2$, we now show photon polarization-dependent band dispersions along the high-symmetry lines (Figs. 3a-d). Overall, the bands around the Fermi energy are mainly derived from the Fe-*d* derived bands shown from the partial density of states (Fig. S3) and depending on the symmetry of the *d*-orbital with respect to the mirror plane (Fig. 2(b)), polarization dependent ARPES intensity is expected. Thus, the distinct polarization dependence enables us to elucidate the orbital characteristics of the main bands. The lambda(Λ)-shaped dispersive bands around the Γ point (ξ band) and the A point (τ band) are predominantly manifested with LV polarization, suggesting that the ξ and τ bands are primarily composed of the $d_{xy}$ and $d_{yz}$ orbitals (odd for $M_{xz}$) defined according to the measurement geometry (Fig. 2b), as supported by the calculated orbital-resolved band dispersion (Figs. 3 e-f and Figs. S4 in SI). Conversely, the low-dispersive band (ε band) is predominantly observed with LH polarizations primarily composed of the $d_{xz}$, $d_{z^2}$, and $d_{x^2-y^2}$ orbitals (even for $M_{xz}$), indicating that the ε band is mainly originated from the $d_{z^2}$ orbital based on the calculated orbital-resolved band dispersion (Figs. 3 e-f and Figs S4 in SI). To provide a clearer visualization, we present the δ, α, and β bands near the Fermi level through momentum distribution curves (MDCs) at $E-E_F$ = -50 meV right above the experimental band dispersions in Fig. 3a-d. The MDCs unveil electron pockets (δ band) at K and H points, along with multiple hole bands (α and β) in all high-symmetry lines, consistent with the results of the DFT calculation (the rightmost panels in Figs. 3 a-d).

Additionally, we observe variations in the band dispersions corresponding to different photon energies (see Figs. S2 in SI for the details).

Compared with $Fe_3GeTe_2$[2,3,5,25], we find that the major features of the band structures, the hole and electron pockets at around Γ-A and K-H lines, respectively, are shared between $Fe_3GaTe_2$ and of $Fe_3GeTe_2$[2,3] except the position of chemical potential. At the Fermi level, there is a decrease in the size of the electron pockets around the Brillouin zone boundary, while the size of the hole pockets around the Γ-point increases, as we move from Ge to Ga. This observation leads us to speculate that the electronic structure of $Fe_3GaTe_2$ closely resembles that of $Fe_3GeTe_2$ doped with one hole/f.u.[2]. This speculation is consistent with that Ga has one less electron than Ge. However, the big increase of $T_C$ in $Fe_3GaTe_2$ contradicts the previous report of the decreased $T_C$ in $Fe_3GeTe_2$ upon hole doping[2]. This discrepancy suggests that the origin of the large increase of $T_C$ in $Fe_3GaTe_2$ system goes beyond the simple changes in chemical potential and electronic structure from Ga substitution. We, therefore, further investigate the relationship between the electronic structures and $T_C$ by obtaining Heisenberg exchange energies ($J_{ex}$)[26] and MAE, based on the accordance between the theoretical and experimental band dispersions. Moreover, as the magnetic properties of $Fe_3GeTe_2$, such as $T_C$, MAE, and coercive field, are reported to be sensitive to doping, we present the dependence of the magnetic properties of $Fe_3GaTe_2$ on the chemical potential shift. This is of importance in potential device applications since the weak vdW interaction between layers allows changing the chemical potential by gating for thin film $Fe_3GaTe_2$[13,27–29].

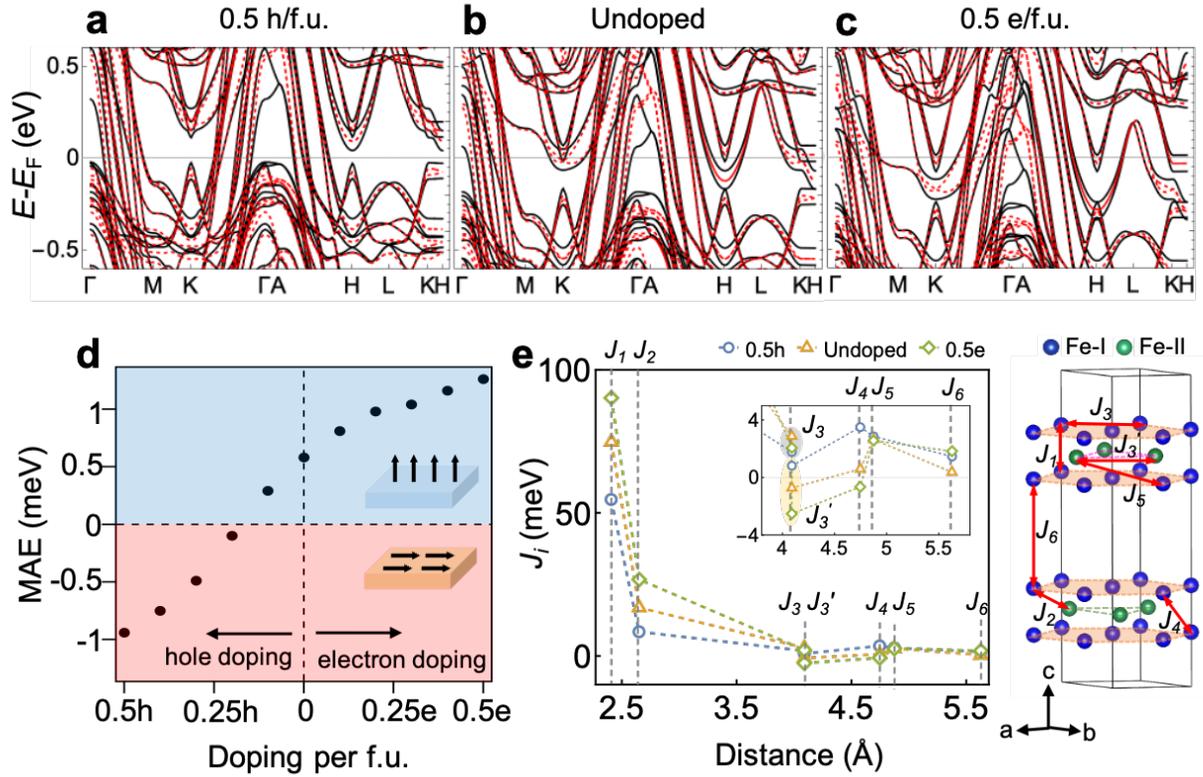

**Figure 4. Doping-dependent electronic and magnetic properties of $Fe_3GaTe_2$ calculated by DFT.** (a-c) DFT band structures with (a) 0.5h doped, (b) undoped, and (c) 0.5e doped $Fe_3GaTe_2$. The doping value is defined by the number of electrons added or removed per formula unit. The black solid lines and red dashed lines are band structures for the out-of-plane and in-plane magnetization, respectively. (d) Magnetocrystalline anisotropy energy (MAE) per Fe atom as a function of doping. The positive and negative signs represent perpendicular and in-plane magnetic anisotropy, respectively. (e) Heisenberg exchange parameters $J_i$ are shown as a function of distances for the 0.5h/f.u.-doped, undoped, and 0.5e/f.u.-doped $Fe_3GaTe_2$. The exchange parameters are defined in the figure on the right, where only Fe-I (blue) and Fe-II (green) atoms are shown with red arrows denoting the pair of Fe atoms corresponding to each exchange parameter. The shaded planes perpendicular to the $c$-axis are formed by connecting Fe-I atoms.

Fig. 4 shows the dependence of the band structures with respect to three representative doping cases (0.5h/f.u., undoped, 0.5e/f.u.). In all cases, there is substantial spin-orbit-coupling driven band splitting where the degeneracy of the bands with in-plane magnetization (red dashed lines) is lifted with out-of-plane magnetization (black solid lines). For the undoped case, there are hole pockets around the Γ and A points ($\alpha$, $\gamma$, $\eta$, and $\beta$ bands) and electron pockets around the K and H points ($\delta$ bands), which are enlarged and reduced with hole and electron doping, respectively. Particularly, we have discovered that the low-dispersive bands along the K-H high symmetry line make significant contributions to the perpendicular magnetocrystalline anisotropy (PMA)[2]. These low-dispersive bands along the K-H line do not cross the Fermi energy for hole doping, whereas the number of low-dispersive bands crossing the Fermi energy along the K-H line increases for electron doping. This suggests a strong dependence of the magnetocrystalline anisotropy energy (MAE) on doping.

Fig. 4(d) presents the MAE as a function of doping in which the MAE of the undoped $Fe_3GaTe_2$ (0.55 meV/Fe or $2.3\times10^6$ J/m$^3$) is close to the experimental value ($4.79\times10^5$ J/m$^3$) at 300 K[19]. More interestingly, we find the sign change in MAE around 0.2h/f.u., suggesting the control of magnetization direction by gating. This switching in magnetization direction could lead to an interesting change in transport properties. Since the anomalous Hall effect (AHE) in the planar geometry requires out-of-plane magnetization and the majority of the AHE is contributed from the splitting of the nodal line connecting K and H points[3], control of AHE by external gating is expected.

In addition to controlling AHE by changing the magnetization direction, we expect a large modulation of the Curie temperature with respect to doping. Fig. 4 (e) presents the Heisenberg exchange parameters for the undoped as well as hole- and electron-doped $Fe_3GaTe_2$, consistent with the previously reported exchange parameters[21]. We find that for the undoped case, the nearest neighbor exchange interaction ($J_1$) is 75 meV about 1.9 times larger than that of $Fe_3GeTe_2$ (40 meV)[30] that accounts for the increase in $T_C$ of 380K compared with $Fe_3GeTe_2$ of 220-230K[21,31,32].

The relatively larger $J_1$ exchange parameter of Fe$_3$GaTe$_2$ could be induced by the structural difference associated with the decreased $c$-lattice constant of Fe$_3$GaTe$_2$, about 1.5% smaller than that of Fe$_3$GeTe$_2$[33–36], where the distance between two Fe-I ions corresponding to the $J_1$ parameter decreases by 2.8%. This is further supported by orbital decomposition of the $J_1$ exchange interaction dominated by the Fe-$d$ orbitals extended along the out-of-plane direction, consisting of ferromagnetic interaction between Fe-$d_{z^2}$ orbitals (43 meV) and between Fe-$d_{xz}/d_{yz}$ orbitals (39 meV) with small antiferromagnetic interaction between $d_{xy}/d_{x^2-y^2}$ orbitals (-7 meV).

With doping the exchange parameters ($J_1$ and $J_2$) mainly contributing to stabilizing the ferromagnetic ordering increase with respect to electron doping and vice versa for hole doping. Combined with MAE dependence, we expect a substantial modulation in $T_C$ by doping and even vanishing $T_C$ for an ultrathin film with easy plane anisotropy by destroying the long-range order[37]. The large increase in the exchange parameters suggests that the substitution of Ge to Ga cannot be simply considered as the change in the electron number but rather induces a substantial change in the magnetic properties leading to much higher $T_C$[38].

In summary, we systematically investigated the electronic structure of Fe$_3$GaTe$_2$ single crystal in its ferromagnetic ground state using ARPES and DFT calculations. We observe qualitative similarities in electronic structure between Fe$_3$GaTe$_2$ and Fe$_3$GeTe$_2$, while significant energy level differences also stand out. Despite substantial variations in ferromagnetic transition temperatures, minor changes in electronic structure and chemical potential exist, necessitating the consideration of other critical parameters. Therefore, our research elucidates the noteworthy contributions of the Heisenberg exchange interaction ($J_{ex}$) and magnetic anisotropy energy (MAE) in the formation of the high Curie temperature ($T_C$) in Fe$_3$GaTe$_2$. Based on the DFT calculations, we anticipate that the magnetic properties can be manipulated through doping, as substantial alterations in MAE and $J_{ex}$ are expected to occur. Our results suggest the comprehensive mechanism of high $T_C$ ferromagnetic order in Fe$_3$GaTe$_2$ and the potential for employing doping strategies to control magnetic interactions in 2D vdW magnets.

**Methods**

**Single crystal growth.** Single crystals of Fe$_3$GaTe$_2$ were grown by the self-flux method. Flakes of Fe (99.98 % purity), Ga (99.99 % purity), and Te (99.99 %) in a molar ratio of 1: 1: 2 were put into a quartz tube. The tube was evacuated and sealed at 0.01 Pa. The sealed quartz ampoule was heated to 1273 K for 10 hours and held there for another day. Then the temperature was quickly decreased down to 1153 K within 2 hours, followed by slow cooling down to 1053 K within 100 hours. Finally, the ampoule was taken out from the furnace and decanted with a centrifuge to separate Fe$_3$GaTe$_2$ single crystals from the flux. The Fe$_3$GaTe$_2$ single crystals were stored in an Ar-filled glovebox to avoid potential degradation.

**ARPES measurements.** Fe$_3$GaTe$_2$ single crystals were cleaved *in situ* in a vacuum of 5 x 10$^{-11}$ Torr. High-resolution ARPES measurements were performed at Beamline 4.0.3 at Advanced Light Source with a sample temperature of 10 K. The energy and angular resolution were set to be ~20 meV and 0.1 degrees, respectively.

**First-principles calculations.** The first-principles DFT calculations were performed with the Vienna Ab initio Simulation Package (VASP)[39,40]. For the exchange-correlation functional, the generalized gradient approximation (GGA) with Perdew-Burke-Ernzerhof (PBE) parameterization[41] was used. The DFT-D2 method of Grimme[42] was used to include the van der Waals interaction. The choice of the exchange-correlation functional gives the lattice constants in better agreement with experimental data, compared to the PBE and PBE with DFT-D3 vdW correction. (See Table. S2 in SI for the details.) The projector augmented wave method[43] was used with an energy cutoff of 600 eV. The Γ-centered 16 × 16 × 5 k-point grid was used. The experimental lattice constants were used with the internal atomic coordinates relaxed with a force threshold of 5 meV/Å. Spin-orbit coupling was included. The force theorem[44,45] was used to calculate the MAE. Electron (hole) doping simulations were treated by increasing (decreasing) total electrons with compensating uniform background charge. The magnetic exchange parameters are calculated by the Green's function method as implemented in the TB2J package[46] within a 15 x 15 x 3 supercell. In the exchange parameter calculations, the Hamiltonian in the atomic orbital basis is extracted using SIESTA code[47] with norm-conserving pseudopotentials and localized pseudoatomic orbitals. The *k*-point mesh of 64x64x16 and the real-space mesh cutoff of 500 Ry are used in the SIESTA calculation.

## ASSOCIATED CONTENT

**Supporting Information.**

$k_z$ dispersion of $Fe_3GaTe_2$; Strong photon energy and photon polarization of $Fe_3GaTe_2$; Partial density of states of $Fe_3GaTe_2$; Calculation of orbital-resolved band structure of $Fe_3GaTe_2$ along the high symmetry lines in the $k_z=\pi/c$ plane; Exchange parameters from the Heisenberg model

## AUTHOR INFORMATION

**Corresponding Authors**

**Hechang Lei** - Beiing Key Laboratory of Optoelectronic Functional Materials MicroNano Devices, Department of Physics, Renmin University of China, Beijing 100872, China and Key Laboratory of Quantum State Construction and Manipulation (Ministry of Education), Renmin University of China, Beijing, 100872, China; Email: hlei@ruc.edu.cn

**Sung-Kwan Mo** - Advanced Light Source, Lawrence Berkeley National Laboratory, Berkeley, CA 94720, USA; Email: skmo@lbl.gov

**Se Young Park** - Department of Physics and Origin of Matter and Evolution of Galaxies (OMEG) Institute, Soongsil University, Seoul 06978, Korea; Email: sp2829@ssu.ac.kr

**Hyejin Ryu** - Center for Spintronics, Korea Institute of Science and Technology (KIST), Seoul 02792, Korea; Email: hryu@kist.re.kr

**Author Contributions**

**J. -E. L. and S. Y. contributed equally to this work.**

**Author Contributions**

H. L., S. Y. P., S.-K. M., and H. R. proposed and designed the research. S. Y. and H. L. performed single-crystal growth. J.-E. L. carried out the ARPES measurements and analyzed the data with assistance from J. H., J. D. D., C. H., S.-K. M. and H. R.; S. O. and S. Y. P. carried out the density functional calculations and provided theoretical support. J.-E. L., S. Y., H. L., S. Y. P., S.-K. M., and H. R. wrote the manuscript and revised it. All authors contributed to the scientific planning and discussions.

**Notes**

The authors declare no competing financial interest.

**ACKNOWLEDGMENT**

The work at the ALS is supported by the US DOE, Office of Basic Energy Sciences, under contract No. DE-AC02-05CH11231. J.-E. L. was supported in part by an ALS Collaborative Postdoctoral Fellowship. Max Planck POSTECH/Korea Research Initiative is supported by the NRF of Korea (2022M3H4A1A04074153). H. C. L. was supported by the National Key R&D Program of China (Grant No. 2018YFE0202600 and 2022YFA1403800), the Beijing Natural Science Foundation (Grant No. Z200005), the National Natural Science Foundation of China (Grants No. 12174443), and the Beijing National Laboratory for Condensed Matter Physics. H. R. acknowledges the KIST Institutional Program (2E32251, 2E32252) and the NRF of Korea grant (No. 2021R1A2C2014179, 2020R1A5A1016518, 2021M3H4A1A03054856). S. Y. P. was supported by the National Research Foundation of Korea (NRF) grant funded by the Korean government (MSIT) (No. 2021R1C1C1009494) and by Basic Science Research Program through the National Research Foundation of Korea (NRF) funded by the Ministry of Education (No. 2021R1A6A1A03043957). C. H. acknowledges support from the National Research Foundation of Korea (NRF) grant funded by the Ministry of Science and ICT (No. 2021R1A2C1004266).